\documentclass[reprint,aip,apl,amsmath,amssymb]{revtex4-1}
\usepackage{amsmath}
\usepackage{amsfonts}
\usepackage{amssymb}
\usepackage{graphics}
\usepackage{float}
\usepackage{graphicx}
\usepackage{epstopdf}
\usepackage[compact]{titlesec}
\usepackage{natbib}
\usepackage{threeparttable}
\usepackage[titletoc,title]{appendix}
\usepackage{lipsum}
\usepackage{varwidth}
\usepackage{tabularx}

\begin{document}
\title{Rashba-driven anomalous Nernst conductivity of lead chalcogenide films}
\author{Parijat Sengupta}
\affiliation{Dept. of Electrical Engineering, University of Illinois, Chicago, IL 60607.}
\author{Junxia Shi}
\affiliation{Dept. of Electrical Engineering, University of Illinois, Chicago, IL 60607.}

\begin{abstract}
{The presence of a finite Berry curvature $\left(\Omega\left(k\right)\right)$ leads to anomalous thermal effects. In this letter, we compute the coefficients for the anomalous Nernst effect $\left(ANE\right) $ and its spin analogue, the spin Nernst effect $\left(SNE\right)$ in lead chalcogenide (\textit{PbX}; \textit{X} = S, Se, Te) films. The narrow gapped \textit{PbX} films with a large spin-orbit coupling (\textit{soc}) offer a significant Rashba interaction that gives rise to $ \Omega\left(k\right) $ and the attendant anomalous thermal behaviour. In presence of a temperature gradient, the $ ANE $ and $ SNE $ establish a thermal and spin current and are characterized by their respective coefficients which acquire higher values for a stronger Rashba interaction. We further show that an extrinsic \textit{soc} generated by an in-plane electric field offers a gate-like mechanism to control (and turn-off) the anomalous thermal currents. Finally, we conclude by deriving the efficiency of an $ ANE $-driven low-temperature Carnot heat engine and demonstrate that it can be gainfully optimized in systems with a robust intrinsic \textit{soc} resulting in low carrier effective masses.}
\end{abstract}
\maketitle

The Nernst effect $\left(NE\right)$ describes the generation of a transverse electric field by a longitudinal temperature gradient in presence of an out-of-plane magnetic field. The related Nernst coefficient is nominally expressed as $ \mathcal{N} = E_{y}/\left(-\nabla T_{x}\right) $.~\cite{abrikosov2017fundamentals} The Nernst-induced electric field and the temperature gradient exist along the \textit{y}- and \textit{x}-axes, respectively. A variant of NE in absence of a real-space magnetic field has also been observed~\cite{xiao2006berry}; the magnetic field, instead, is supplied by an analogous quantity - the Berry curvature.~\cite{gradhand2012first} The Berry curvature $\left(\Omega\left(k\right)\right)$ as an effective magnetic field in momentum space $\left(k\right)$ imparts a Lorentz force on carrier electrons giving rise to an `anomalous' velocity that is the precursor to a variety of observed effects, an illustration of which is the anomalous Nernst effect $\left(ANE\right)$. The $ ANE $ has been theoretically predicted in a wide selection of materials including the \textit{d}-density wave state in cuprate superconductors~\cite{zhang2008anomalous}, illuminated graphene~\cite{zhou2015anomalous}, and monolayer group-VI dichalcogenides.~\cite{yu2015thermally} In each of these family of materials, it is possible to write a prototypical Hamiltonian transformable along a closed contour in momentum space in a cyclic adiabatic process giving rise to the non-zero Berry connection $\left(\mathcal{A}\left(k\right)\right)$ from which an equivalent magnetic field $\left(\Omega\left(k\right) = \nabla_{k} \times \mathcal{A}\left(k\right)\right)$ can be defined. For the case of graphene-like materials and the dichalcogenides, in the presence of either broken inversion or time-reversal symmetry (\textit{TRS}), the form of Hamiltonian that sets up a finite $ \Omega\left(k\right) $ is of a massive Dirac-type : $ \mathcal{H}_{eff} = \nu\left(\sigma_{x}k_{y} - \sigma_{y}k_{x}\right) + \Delta\sigma_{z} $. The constant $ \nu $ has units of $ eV\AA $ and $ \Delta $ is the generalized Dirac mass. The Pauli matrices $\left(\sigma_{i}\right)$ may act on the lattice or spin sub-space.

It is easy to note, however, that the form of the $ k $-dependent part of the Hamiltonian that permits a finite $ \Omega\left(k\right) $ also describes the linear Rashba spin-orbit coupling $\left(RSOC\right)$ for Bloch conduction electrons.~\cite{averkiev2002spin} For a set of conduction electrons in a thin film (quantum well) that follow a quadratic dispersion and split into the linear \textit{RSOC}-induced spin-polarized sub-bands, it is reasonable to anticipate the occurrence of a similarly definable $ \Omega\left(k\right) $. The $ \Omega\left(k\right) $ in this case would be solely an outcome of the linear \textit{RSOC} Hamiltonian; the quadratic term does not contribute. A non-vanishing $ \Omega\left(k\right) $ therefore alludes to the appearance of a concomitant $ ANE $, the analytic estimation of which is the chief purpose here. We quantitatively estimate the strength of $ ANE $ in thin films whose Bloch conduction bands are split by RSOC in spin-polarized ensembles and examine underlying dependencies that enhance this thermomagnetic process. An additional purported aim is also to uncover avenues that potentially optimize the efficiency of $ ANE $ via changes to strength of \textit{RSOC}, the band curvature (tied to film dimensions), and external impurities. In fact, since an avowed goal in recent times hinges on the design of `energy-harvesting' techniques, a large $ ANE $ can complement the $ NE $ in miniaturized magneto-thermal devices. 

It is worthwhile though to clarify that while a definite $ \Omega\left(k\right) $ is derivable from a \textit{RSOC} Hamiltonian and assumes an identical form to that obtained for gapped graphene-like materials and chalcogenides~\cite{xiao2012coupled}, the genesis of it lies in the spin degree of freedom unlike an inversion breaking mixing of orbitals in the latter. This disparity in the origin of the emergence of $ \Omega\left(k\right) $ aside, it is significant to observe that \textit{RSOC} is only operational when inversion symmetry is lost, which essentially constitutes one of the prerequisites (the other is \textit{TRS} and at least one must be satisfied) for a non-vanishing $ \Omega\left(k\right) $ and fulfilled by the graphene family and dichalcogenides. Evidently, for a discernible $\Omega\left(k\right) $ (and $ ANE $), a primary requirement centers around a large \textit{RSOC}, a quantity generally pronounced in confined structures of compounds with narrow band gaps and high intrinsic spin-orbit coupling. While several sets of materials display a robust \textit{RSOC}, it is beneficial to recall the thermal basis of the parent \textit{NE} and thus select a candidate system that also combines favourable thermoelectric behaviour. The lead chalcogenides~\cite{ravich2013semiconducting} - \textit{PbX} (\textit{X} = S/Se/Te) -  conform well in this regard, possessing the necessary material attributes for a large \textit{RSOC} and a high thermoelectric figure of merit (\textit{ZT}). PbTe and its alloyed derivatives have been widely researched for achieving an enhanced \textit{ZT}.~\cite{dughaish2002lead}   

\begin{figure}[t!]
`\includegraphics[scale=0.74]{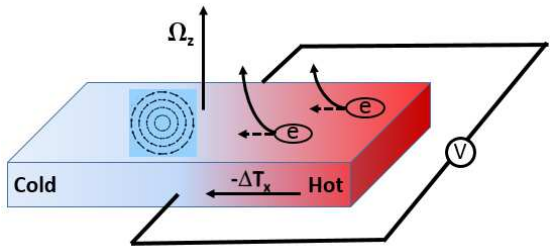}
\vspace{-0.2cm}
\caption{A schematic depiction of the anomalous Nernst effect in a PbTe (representative \textit{PbX}) slab with a temperature gradient along the \textit{x}-axis. The Rashba coupled Bloch conduction electrons with the characteristic helical in-plane spin polarization shown on the slab surface generates $ \Omega\left(k\right) $, a momentum-dependent magnetic field along the \textit{z}-axis (out-of-plane). An external \textit{z}-axis aligned magnetic field (not shown) is also applied to the PbTe slab. An anomalous voltage develops in a direction transverse (\textit{y}-axis) to $ \nabla_{x}T $.}
\label{sch}
\vspace{-0.5cm}
\end{figure}

In an \textit{n}-doped PbTe sample under an out-of-plane magnetic field (causing a Zeeman split) and a temperature gradient (Fig.~\ref{sch}), we show that the $ ANE $ responds to a Rashba-controlled Berry curvature distribution in momentum space and can be further modulated with an in-plane electric field. A complete cessation of $ ANE $ (vanishing $ \Omega\left(k\right) $ happens when the in-plane electric field initiated spin-orbit coupling (\textit{soc}) exactly annuls the Zeeman splitting. An accompanying quantifiable spin current - the anomalous spin Nernst effect - also flows mirroring the pattern observed for $ ANE $. In the last part, we develop the idea of an $ ANE $- driven Carnot engine whose efficiency is shown to be optimized by low carrier effective masses - the hallmark of high \textit{soc} that also greatly influences \textit{RSOC}. 

For an analytic formulation, we begin by writing the expression for $ ANE $ which is $ J_{y} = \mathcal{N}^{'}\left(-\nabla_{x}T\right)$. The primed coefficient $ N^{'} $ distinguishes from $ \mathcal{N} $, the corresponding notation for $ NE $. The $ ANE $ coefficient is~\cite{xu2016detecting}
\begin{equation}
\mathcal{N}^{'} = \dfrac{ek_{B}}{\hbar}\sum\limits_{\pm}\int \dfrac{d^{2}\mathbf{k}}{4\pi^{2}}\,\Omega\left(k\right)\,\mathcal{S}\left(k\right).
\label{anerel}
\end{equation} 
where $ \mathcal{S}\left(k\right) $ is the entropy density. The entropy density is defined as: $ \mathcal{S}\left(k\right) = -f_{k}lnf_{k} - (1 - f_{k})ln(1 - f_{k}) $. Here $ f_{k} $ is the usual Fermi distribution function. The summation over the spin-split bands is indicated by $ \pm $ under the $ \sum $ operator. As a brief insight into the particular form of the $ ANE $ coefficient (Eq.~\ref{anerel}), we simply note that a finite $ \Omega\left(k\right) $ lets the electron carriers (of charge `$e$' and in presence of an electric field $ \mathbf{E} $) acquire an additional velocity $ v_{ane} = \left(e\mathbf{E}/\hbar\right) \times \Omega\left(k\right) $. Multiplying $ v_{ane} $ by the entropy density furnishes the coefficient for the transverse heat current from which we obtain $ \mathcal{N}^{'} $ in Eq.~\ref{anerel}. The primary task, therefore, to proceed further with $ ANE $ calculations is a determination of $ \Omega\left(k\right) $. To do so, we first write down the minimal Hamiltonian that describes the Rashba and Zeeman spin-split parabolic conduction bands in a \textit{PbX} quantum well. It is simply given by
\begin{equation}
H_{0} = \dfrac{p^{2}}{2m^{*}} + \alpha_{R}\left(\sigma_{x}k_{y} - \sigma_{y}k_{x}\right) + \Delta\sigma_{z}.
\label{ham1}
\end{equation}
The Pauli matrices in Eq.~\ref{ham1} act on the spin-space and the parameter $ \Delta $ is the the out-of-plane magnetic field governed Zeeman splitting. The Rashba coupling parameter is $ \alpha_{R} $ with adjustable strength via a gate electrode while the effective mass is $ m^{*} $, which here are for the \textit{L}-valley conduction electrons of a \textit{PbX} film (quantum well confined along the \textit{z}-axis). The corresponding eigen states are $ \varepsilon_{\pm} = p^{2}/2m^{*} \pm \alpha_{R}k \pm \Delta $. The upper (lower) sign is for the spin-up (down) band. Additionally, we consider only conduction electrons (\textit{CE}) of the \textit{PbX} film (rock salt crystal structure) that belong to the \textit{L}-valley and whose axis coincides with the $ \left[111\right] $ direction. Note that there exist three other oblique valleys with axes mis-aligned to the $ \left[111\right] $ vector.~\cite{spl} We make use of a $ 4 \times 4 $ \textit{k.p} Hamiltonian~\cite{dimmock1964band,kang1997electronic} that captures the dispersion around the high-symmetry \textit{L}-valley; substituting for the confined $ k_{z} = -i\partial_{z} $ in the Hamiltonian and followed by a numerical diagonalization supplies the quantum well \textit{CE} effective mass.~\cite{spl} The presented calculations use PbTe as the representative \textit{PbX}. 

For a quantum well, which is a two-dimensional system whose Hamiltonian (disregarding the quadratic component that does not contribute to $ \Omega\left(k\right)$ ) is expressible as $ H\left(k\right) = \mathbf{d\left(k\right)}\cdot \mathbf{\sigma} $, the Berry curvature is defined as~\cite{shen2012topological} 
\begin{equation}
\Omega_{\mu\nu} = \dfrac{1}{2}\varepsilon_{\alpha\beta\gamma}\hat{d}_{\alpha}\left(\mathbf{k}\right)\partial_{k_{\mu}}\hat{d}_{\beta}\left(\mathbf{k}\right)\partial_{k_{\nu}}\hat{d}_{\gamma}\left(\mathbf{k}\right),
\label{berry}
\end{equation}
where $ \hat{\mathbf{d}}\left(\mathbf{k}\right) = \dfrac{\mathbf{d\left(\mathbf{k}\right)}}{d\left(k\right)} $. Applying this formalism in the case of Rashba Hamiltonian (Eq.~\ref{ham1}), the $ \Omega\left(k\right) $ takes the form:
\begin{equation}
\Omega\left(k\right) = \pm\dfrac{\alpha_{R}^{2}\Delta}{2\left[\left(\alpha_{R}k\right)^2 + \Delta^2\right]^{3/2}}\hat{\mathbf{z}}.
\label{bc}
\end{equation}
In Eq.~\ref{bc}, $ k^{2} = k_{x}^{2} + k_{y}^{2} $. The upper (lower) sign is for the spin-down (up) band. The $ \Omega $ as a momentum-dependent magnetic field points out-of-plane (the \textit{z}-axis). The $ \Omega\left(k\right) $ vanishes as $ \Delta \rightarrow 0 $; the quenching of $ \Delta $ (the Zeeman splitting) in this case restores \textit{TRS} from which follows the relation, $\Omega\left(k\right) = 0 $. A simple inspection of Eqs.~\ref{anerel} and ~\ref{bc} reveals that for a significant $ ANE $ a large $ \Omega\left(k\right) $ is desirable which in turn requires a sizable Rashba splitting determined via the strength of the coefficient, $ \alpha_{R} $. The Rashba coefficient is strong in narrow band gap materials with considerable intrinsic \textit{soc}, such as those belonging to the \textit{PbX} family. The Rashba coupling coefficient is expressed as : $ \alpha_{R} = \lambda_{0}\langle\,F\left(z\right)\rangle $, where $ \langle\,F\left(z\right)\rangle $ is the average out-of-plane (\textit{z}-axis) electric field. The average value for $ \langle\,F\left(z\right)\rangle $ is $ en/\epsilon $. Here, $ e $ is the electronic charge, the dopant density is $ n $, and $ \epsilon $ identifies the dielectric constant. The material-dependent $ \lambda_{0} $ is given as~\cite{e1997spin}
\begin{equation}
\lambda_{0} = \dfrac{\hbar^{2}}{2m^{*}}\dfrac{\Delta_{so}}{E_{g}}\dfrac{2E_{g}+\Delta_{so}}{\left(E_{g} + \Delta_{so}\right)\left(3E_{g} + 2\Delta_{so}\right)}.
\label{rasz}
\end{equation}
For the specific case of \textit{PbX} quantum wells, the parameters in Eq.~\ref{rasz} are defined at the \textit{L}-valley; here, $ E_{g} $ is the direct band gap, the intrinsic \textit{soc} is $ \Delta_{so} $, and $ m^{*} $ denotes the conduction band effective mass. The tuning of $ \alpha_{R} $ is therefore, unlike, the intrinsic \textit{soc} possible via changes to the band gap and effective mass in confined structures.

Before we carry out a quantitative analysis of the anomalous thermal behaviour, a set of remarks are in order: Firstly (1), the dielectric constant (\textit{dc}) of PbTe is abnormally large $\left(\approx\, 400\right)$, an outcome attributed to the high-polarizability of the chemical bond. This high \textit{dc}~\cite{kanai1963dielectric} in addition to determining $ \alpha_{R} $ via $ \langle\,F\left(z\right)\rangle $ also couples with the low effective electron masses (in part, attributed to a substantial intrinsic \textit{soc}) to set up a significant Bohr radius~\cite{heremans2017tetradymites} and thus enhancing carrier mobility. While thermoelectric applications require a pronounced mobility (and low thermal conductivity) for an optimized thermoelectric figure-of-merit (\textit{ZT}), for the Rashba-driven $ ANE $, a change in \textit{dc} is reflected in $ \alpha_{R} $ which clearly revises $ \Omega\left(k\right) $ and consequently the thermal anomalous effects. The \textit{dc} has been shown~\cite{alves2013lattice} to be adjustable via simple lattice deformations of the rock salt crystal. In passing, it is useful to mention that polarizable \textit{PbX} bonds also typically scupper the thermal conductivity to improve \textit{ZT}. The second comment (2) pertains to additional \textit{soc} terms that may occur in the Hamiltonian (Eq.~\ref{ham1}). An extra \textit{soc}-term (besides Rashba) for an in-plane electric field $\left(F_{ip}\right)$ can be of the form $ e\beta\hat{\sigma}\cdot\left(\mathbf{E}\times\mathbf{k}\right)$. For an \textit{x}-axis directed $ F_{ip} $, the Hamiltonian receives a contribution expressed as $ e\beta F_{ip}k_{y}\sigma_{z} $. The corresponding expression for $\Omega\left(k\right)$ by a direct application of the formula in Eq.~\ref{berry} gives
\begin{equation}
\Omega\left(k\right) = \pm\dfrac{\alpha_{R}^{2}\left(\Delta + e\beta F_{ip}k_{y}\right)}{2\left[\left(\alpha_{R}k\right)^2 + \left(\Delta + e\beta F_{ip}k_{y}\right)^2\right]^{3/2}}\hat{\mathbf{z}}.
\label{nbc}
\end{equation}
A more appealing situation emerges for an in-plane electric field solely directed along the \textit{y}-axis; the \textit{soc} in this case is simply $ -e\beta F_{ip}k_{x}\sigma_{z} $ and manifestly counteracts $ \Delta $, the Zeeman splitting. For values of the \textit{y}-directed $ F_{ip} $ such that $ \Delta - e\beta F_{ip}k_{x} \rightarrow 0 $, the out-of-plane (\textit{z}-axis) magnetic field induced broken \textit{TRS} is restored. The fulfillment of \textit{TRS} to which we pointed out before leads to a ceasing of $\Omega\left(k\right) $ and the attendant $ ANE $. It is therefore also apparent (from Eq.~\ref{nbc}) that a union of the spin-orbit Hamiltonians through their respective coupling coefficients, $ \alpha_{R} $ and $ \beta $, allows a more detailed measure of control over the $ ANE $-governed charge current (in a closed circuit). In fact, $ \mathcal{F}_{ip} $ can be considered applied from a gate terminal and serve as a threshold bias; for the correct polarity and magnitude, as $ \Omega\left(k\right) \rightarrow 0 $, it describes a complete turnoff setting unique to the material system. The final remark (3) considers the overall contribution of the two spin-split bands. Noting that $\Omega_{\downarrow}\left(k\right) = - \Omega_{\uparrow}\left(k\right) $, the complete $ ANE $ coefficient becomes $ \mathcal{N}^{'}_{ov} = ek_{B}/\left(4\pi^{2}\hbar\right)\int\mathbf{d}^{2}k\Omega_{\downarrow}\left(k\right)\left[\mathcal{S}_{\downarrow}\left(k\right) - \mathcal{S}_{\uparrow}\left(k\right)\right] $. It is therefore straightforward to see that a spin-up band placed energetically above its spin-down counterpart when empty (or zero entropy) maximizes the $ ANE $. Further, in connection to the spin-split bands and analogous to $ ANE $, following Ref.~\onlinecite{yu2015thermally}, a spin Nernst coefficient $\left(SNE\right)$ can be defined as
\begin{equation}
\mathcal{N}^{'}_{s} = \dfrac{k_{B}}{2}\int\dfrac{\mathbf{d}^{2}k}{4\pi^{2}}\left[\Omega_{\uparrow}\left(k\right)\mathcal{S}_{\uparrow}\left(k\right) - \Omega_{\downarrow}\left(k\right)\mathcal{S}_{\downarrow}\left(k\right)\right].
\label{sne}
\end{equation}
It is, however, useful to reiterate that the $ \Omega\left(k\right) $ in Ref.~\onlinecite{yu2015thermally} strictly arises from the broken inversion symmetry of the monolayer transition metal dichalcogenide, unlike the Rashba-governed case here. 

\begin{figure}
\includegraphics[scale=0.69]{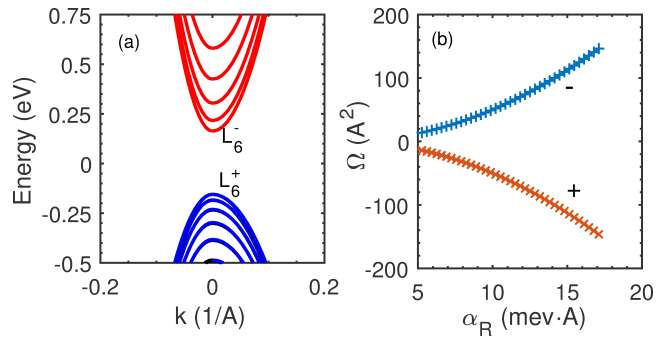}
\vspace{-0.3cm}
\caption{The numerically obtained $ L $-valley dispersion of a $ 6.0\,nm $ wide $ \left[111\right] $ PbTe film along the high-symmetry path $ \overline{K}-\overline{L}-\overline{\Gamma} $ is shown on the left panel (a). The right figure (b) plots (using Eq.~\ref{bc}) the Rashba-aided Berry curvature at the conduction band minimum $\left(\vert k \vert = 0\right)$ where the upper (lower) branch is for the spin-down (up) conduction state. The asymmetry-inducing electric field (out-of-plane) necessary for the Rashba splitting arises from an \textit{n}-doping concentration; for purpose of numerical calculation, $ n $ was varied between $ 1 \times 10^{12}\, cm^{-2} $ and $ 4 \times 10^{12}\,cm^{-2} $. The dielectric constant of PbTe was set to 400 (see note below). The Zeeman splitting is treated as an external parameter and set to $ \Delta = 2.0\,meV $ throughout.}
\label{dber}
\vspace{-0.73cm}
\end{figure}

\begin{figure}[t!]
\includegraphics[scale=0.72]{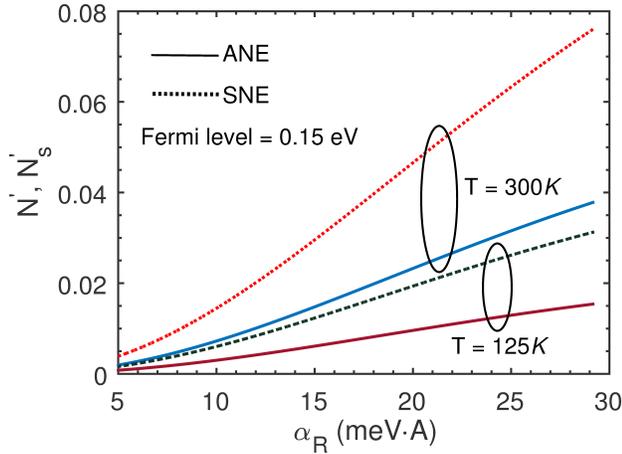}
\vspace{-0.2cm}
\caption{The $ ANE \left(\mathcal{N}^{'}\right) $ and $ SNE \left(\mathcal{N}^{'}_{s}\right) $ coefficients are plotted in units of $ ek_{B}/h $ and $ k_{B}/4\pi $, respectively. The $\mathcal{N}^{'}$ is shown only for the spin-down band. Both coefficients increase as higher values of $ \alpha_{R} $ are realized through doping or an external gate electrode. Additionally, at elevated temperatures that raise the entropy, larger coefficients are obtained. While the overall $ \mathcal{N}^{'} $ taking both spin split bands into account nearly vanishes, the $ SNE $ allows the flow of a net spin current.}
\label{asnc}
\vspace{-0.68cm}
\end{figure}

For numerical estimate of $ \alpha_{R} $, from which follows the $ \Omega\left(k\right) $ and coefficients for $ ANE $ and $ SNE $, a $ 6.0\, nm $ wide PbTe film grown along the $ \left[111\right] $ axis is selected as the model structure. The \textit{L}-valley band gap and effective mass (transverse) of this film from a \textit{k.p} calculation are $ 0.0565m_{0} $ and $ 0.33\, eV $. The free electron mass is $ m_{0} = 9.1 \times 10^{-31}\, kg $. The dispersion of the $ 6.0\, nm $ wide PbTe film and the accompanying Rashba-induced $ \Omega\left(k\right) $ is shown in Fig.~\ref{dber}. Note that $ \Omega\left(k\right) $ is plotted as a function of $ \alpha_{R} $, which is a function of material parameters and the film's Bloch conduction electrons effective mass.
To proceed further a number of other parameters useful in determination of $ \mathcal{N}^{'} $ and $ \mathcal{N}_{s}^{'} $ must be defined: We begin by assigning the temperature $\left(T\right)$ a pair of values: $ T = \lbrace 125, 300\rbrace K $. The Fermi level is set to $ E_{f} = 0.15\, eV $ from the bottom of the conduction band while the charge/dopant density is assumed to lie between $ 10^{12}\,cm^{-2} $ and $ 8 \times 10^{12}\,cm^{-2} $. This dopant density furnished electric field lets $ \alpha_{R} $ acquire values from $ 5.0\,meV\AA - 30.0\,meV\AA $. Inserting these numbers in Eqs.~\ref{anerel} and ~\ref{sne} and numerically integrating for $ \vert k \vert \leq 0.3\,1/\AA $, we plot $ \mathcal{N}^{'} $ and $ \mathcal{N}^{'}_{s} $ in Fig.~\ref{asnc} in units of $ ek_{B}/h $ and $ k_{B}/4\pi $, respectively. We only show the $ ANE $ coefficient for the spin-down band since the contribution of the spin-up band differs marginally from the former and carries a reversed sign. The closeness is simply a consequence of the moderate energy difference between the spin-split bands. Separately, the plot clearly reveals a more forceful display of $ ANE $ and $ SNE $ for weightier $ \alpha_{R} $, which expressly influences and enlarges $ \Omega\left(k\right) $ - the engine behind anomalous effects. We make a note here that the parameter $ \alpha_{R} $, in addition to dopant density changes is also amenable to further modification via adjustments to $ m^{*} $, the band gap $\left(E_{g}\right)$, and the intrinsic \textit{soc}. While the \textit{soc}, admittedly, is harder to modulate; however, $ m^{*} $ and $ E_{g} $ through varying degrees of confinement, layered-heterostructure design, and strain-like perturbation can substantially augment $ \alpha_{R} $. In line with schemes that may reinforce the anomalous thermal behaviour, it is also expedient to identify regions in momentum-space where $ \Omega\left(k\right) $ and $ \mathcal{S}\left(k\right) $ attain their highest values. The $ \Omega\left(k\right) $ from Eq.~\ref{bc} has a Lorentzian spread centered around the $ \vert k \vert = 0 $ point, which is the conduction band origin and reaches its maximum; likewise, the entropy has peaks on the Fermi surface and tails off away from it. For these two variables to amplify $ ANE $ and $ SNE $, an intersecting region of momentum space must therefore be chosen to locate carriers with energy closely aligned to the Fermi surface while simultaneously ensuring that it isn't too far away from $ \vert k \vert = 0 $ for a reasonable $ \Omega\left(k\right)$.

As a more definitive guide that ascertains the efficiency ($\eta $ = output/input) of $ ANE $, we can construct a paradigmatic Carnot engine like abstraction into which heat is pumped and `useful' work extracted as power in a closed circuit. The power (`output') is $ \left(\mathcal{N}^{'}(-\nabla_{x}T\right)^{2}\mathcal{R}$. The electric resistance of the closed circuit is $ \mathcal{R} $. We consider a low temperature regime to ignore any phonon-driven thermal currents. A Carnot engine modeled on the $ NE $ must proceed by establishing a temperature gradient, where the desired heat current (`input') to maintain a temperature difference is given by the Fourier law : $ J_{Q} = -\kappa\nabla_{x}T $. Here, $ \kappa $ is the thermal conductivity which is connected via the Wiedemann-Franz law (WFL) to its electric counterpart.~\cite{sengupta2017low} Briefly, the electric conductivity (for energy $\varepsilon$) using the linearized Boltzmann equation is $ \sigma = e^{2}v_{f}^{2}/2\int\, d\varepsilon D\left(\varepsilon\right)\tau\left(\varepsilon\right)\left(-\partial_{\varepsilon}f\right)$. The density-of-states, $ D\left(\varepsilon\right) $, ignoring the linear Rashba  and Zeeman term is $ m^{*}/\left(\pi\hbar^{2}\right)$ and the Fermi velocity $\left(v_{f}\right) $ is $ \hbar k/m^{*} $. The scattering time is $\tau\left(\varepsilon\right)$. A direct application of WFL therefore gives the thermal conductivity as $ \kappa = \mathcal{L}\sigma T $. Here, $ \mathcal{L} = 2.44 \times 10^{-8}\, W\Omega K^{-2} $ is the Lorentz number. By following the outlined sequence of steps, the quantity $\eta $ for the proposed Carnot engine is 
\begin{equation}
\eta_{Carnot} = \dfrac{2\pi\hbar^{2}\left[N^{'2}\nabla_{x}T\right]\mathcal{R}}{\mathcal{L}e^{2}v_{f}^{2}m^{*}\tau T}.
\label{cef}
\end{equation}
It is clearly noticeable from Eq.~\ref{cef} that a low-effective mass improves efficiency - a result that ties well with the requirement of a strong Rashba coupled material, as both arise in compounds with a large intrinsic \textit{soc}. 

To summarize, we obtained analytic expressions for anomalous Nernst and spin Nernst coefficients tunable through the Rashba-created Berry curvature in PbTe films. Similarly, it is expected that narrow-gap and strongly spin-orbit coupled III-V materials such as InAs or InSb can give rise to comparable anomalous thermal currents. Besides, the $ SNE $-origin anomalous spin current may find applications in spin caloritronics~\cite{bauer2012spin} for a more diverse set of PbTe-like materials, rather than being limited, as it is hitherto, to mostly magnetic systems.


\begin{thebibliography}{21}
\expandafter\ifx\csname natexlab\endcsname\relax\def\natexlab#1{#1}\fi
\expandafter\ifx\csname bibnamefont\endcsname\relax
  \def\bibnamefont#1{#1}\fi
\expandafter\ifx\csname bibfnamefont\endcsname\relax
  \def\bibfnamefont#1{#1}\fi
\expandafter\ifx\csname citenamefont\endcsname\relax
  \def\citenamefont#1{#1}\fi
\expandafter\ifx\csname url\endcsname\relax
  \def\url#1{\texttt{#1}}\fi
\expandafter\ifx\csname urlprefix\endcsname\relax\def\urlprefix{URL }\fi
\providecommand{\bibinfo}[2]{#2}
\providecommand{\eprint}[2][]{\url{#2}}

\bibitem[{\citenamefont{Abrikosov}(2017)}]{abrikosov2017fundamentals}
\bibinfo{author}{\bibfnamefont{A.}~\bibnamefont{Abrikosov}},
  \emph{\bibinfo{title}{Fundamentals of the Theory of Metals}}
  (\bibinfo{publisher}{Courier Dover Publications}, \bibinfo{year}{2017}).

\bibitem[{\citenamefont{Xiao et~al.}(2006)\citenamefont{Xiao, Yao, Fang, and
  Niu}}]{xiao2006berry}
\bibinfo{author}{\bibfnamefont{D.}~\bibnamefont{Xiao}},
  \bibinfo{author}{\bibfnamefont{Y.}~\bibnamefont{Yao}},
  \bibinfo{author}{\bibfnamefont{Z.}~\bibnamefont{Fang}}, \bibnamefont{and}
  \bibinfo{author}{\bibfnamefont{Q.}~\bibnamefont{Niu}},
  \bibinfo{journal}{Physical review letters} \textbf{\bibinfo{volume}{97}},
  \bibinfo{pages}{026603} (\bibinfo{year}{2006}).

\bibitem[{\citenamefont{Gradhand et~al.}(2012)\citenamefont{Gradhand, Fedorov,
  Pientka, Zahn, Mertig, and Gy{\"o}rffy}}]{gradhand2012first}
\bibinfo{author}{\bibfnamefont{M.}~\bibnamefont{Gradhand}},
  \bibinfo{author}{\bibfnamefont{D.}~\bibnamefont{Fedorov}},
  \bibinfo{author}{\bibfnamefont{F.}~\bibnamefont{Pientka}},
  \bibinfo{author}{\bibfnamefont{P.}~\bibnamefont{Zahn}},
  \bibinfo{author}{\bibfnamefont{I.}~\bibnamefont{Mertig}}, \bibnamefont{and}
  \bibinfo{author}{\bibfnamefont{B.}~\bibnamefont{Gy{\"o}rffy}},
  \bibinfo{journal}{Journal of Physics: Condensed Matter}
  \textbf{\bibinfo{volume}{24}}, \bibinfo{pages}{213202}
  (\bibinfo{year}{2012}).

\bibitem[{\citenamefont{Zhang et~al.}(2008)\citenamefont{Zhang, Tewari,
  Yakovenko, and Sarma}}]{zhang2008anomalous}
\bibinfo{author}{\bibfnamefont{C.}~\bibnamefont{Zhang}},
  \bibinfo{author}{\bibfnamefont{S.}~\bibnamefont{Tewari}},
  \bibinfo{author}{\bibfnamefont{V.}~\bibnamefont{Yakovenko}},
  \bibnamefont{and} \bibinfo{author}{\bibfnamefont{S.~D.} \bibnamefont{Sarma}},
  \bibinfo{journal}{Physical Review B} \textbf{\bibinfo{volume}{78}},
  \bibinfo{pages}{174508} (\bibinfo{year}{2008}).

\bibitem[{\citenamefont{Zhou et~al.}(2015)\citenamefont{Zhou, Xu, and
  Jin}}]{zhou2015anomalous}
\bibinfo{author}{\bibfnamefont{X.}~\bibnamefont{Zhou}},
  \bibinfo{author}{\bibfnamefont{Y.}~\bibnamefont{Xu}}, \bibnamefont{and}
  \bibinfo{author}{\bibfnamefont{G.}~\bibnamefont{Jin}},
  \bibinfo{journal}{Physical Review B} \textbf{\bibinfo{volume}{92}},
  \bibinfo{pages}{235436} (\bibinfo{year}{2015}).

\bibitem[{\citenamefont{Yu et~al.}(2015)\citenamefont{Yu, Zhu, Su, and
  Jauho}}]{yu2015thermally}
\bibinfo{author}{\bibfnamefont{X.}~\bibnamefont{Yu}},
  \bibinfo{author}{\bibfnamefont{Z.}~\bibnamefont{Zhu}},
  \bibinfo{author}{\bibfnamefont{G.}~\bibnamefont{Su}}, \bibnamefont{and}
  \bibinfo{author}{\bibfnamefont{A.-P.} \bibnamefont{Jauho}},
  \bibinfo{journal}{Physical review letters} \textbf{\bibinfo{volume}{115}},
  \bibinfo{pages}{246601} (\bibinfo{year}{2015}).

\bibitem[{\citenamefont{Averkiev et~al.}(2002)\citenamefont{Averkiev, Golub,
  and Willander}}]{averkiev2002spin}
\bibinfo{author}{\bibfnamefont{N.}~\bibnamefont{Averkiev}},
  \bibinfo{author}{\bibfnamefont{L.}~\bibnamefont{Golub}}, \bibnamefont{and}
  \bibinfo{author}{\bibfnamefont{M.}~\bibnamefont{Willander}},
  \bibinfo{journal}{Journal of physics: condensed matter}
  \textbf{\bibinfo{volume}{14}}, \bibinfo{pages}{R271} (\bibinfo{year}{2002}).

\bibitem[{\citenamefont{Xiao et~al.}(2012)\citenamefont{Xiao, Liu, Feng, Xu,
  and Yao}}]{xiao2012coupled}
\bibinfo{author}{\bibfnamefont{D.}~\bibnamefont{Xiao}},
  \bibinfo{author}{\bibfnamefont{G.}~\bibnamefont{Liu}},
  \bibinfo{author}{\bibfnamefont{W.}~\bibnamefont{Feng}},
  \bibinfo{author}{\bibfnamefont{X.}~\bibnamefont{Xu}}, \bibnamefont{and}
  \bibinfo{author}{\bibfnamefont{W.}~\bibnamefont{Yao}},
  \bibinfo{journal}{Physical Review Letters} \textbf{\bibinfo{volume}{108}},
  \bibinfo{pages}{196802} (\bibinfo{year}{2012}).

\bibitem[{\citenamefont{Ravich}(2013)}]{ravich2013semiconducting}
\bibinfo{author}{\bibfnamefont{I.~I.} \bibnamefont{Ravich}},
  \emph{\bibinfo{title}{Semiconducting lead chalcogenides}},
  vol.~\bibinfo{volume}{5} (\bibinfo{publisher}{Springer Science \& Business
  Media}, \bibinfo{year}{2013}).

\bibitem[{\citenamefont{Dughaish}(2002)}]{dughaish2002lead}
\bibinfo{author}{\bibfnamefont{Z.}~\bibnamefont{Dughaish}},
  \bibinfo{journal}{Physica B: Condensed Matter}
  \textbf{\bibinfo{volume}{322}}, \bibinfo{pages}{205} (\bibinfo{year}{2002}).

\bibitem[{\citenamefont{Xu et~al.}(2016)\citenamefont{Xu, Zhou, and
  Jin}}]{xu2016detecting}
\bibinfo{author}{\bibfnamefont{Y.}~\bibnamefont{Xu}},
  \bibinfo{author}{\bibfnamefont{X.}~\bibnamefont{Zhou}}, \bibnamefont{and}
  \bibinfo{author}{\bibfnamefont{G.}~\bibnamefont{Jin}},
  \bibinfo{journal}{Applied Physics Letters} \textbf{\bibinfo{volume}{108}},
  \bibinfo{pages}{203104} (\bibinfo{year}{2016}).

\bibitem[{\citenamefont{Dimmock and Wright}(1964)}]{dimmock1964band}
\bibinfo{author}{\bibfnamefont{J.}~\bibnamefont{Dimmock}} \bibnamefont{and}
  \bibinfo{author}{\bibfnamefont{G.}~\bibnamefont{Wright}},
  \bibinfo{journal}{Physical Review} \textbf{\bibinfo{volume}{135}},
  \bibinfo{pages}{A821} (\bibinfo{year}{1964}).

\bibitem[{\citenamefont{Kang and Wise}(1997)}]{kang1997electronic}
\bibinfo{author}{\bibfnamefont{I.}~\bibnamefont{Kang}} \bibnamefont{and}
  \bibinfo{author}{\bibfnamefont{F.~W.} \bibnamefont{Wise}},
  \bibinfo{journal}{JOSA B} \textbf{\bibinfo{volume}{14}},
  \bibinfo{pages}{1632} (\bibinfo{year}{1997}).

\bibitem[{spl()}]{spl}
\bibinfo{note}{See accompanying note in Supplementary Material}.

\bibitem[{\citenamefont{Shen}(2012)}]{shen2012topological}
\bibinfo{author}{\bibfnamefont{S.-Q.} \bibnamefont{Shen}},
  \emph{\bibinfo{title}{Topological insulators}}, vol. \bibinfo{volume}{174}
  (\bibinfo{publisher}{Springer}, \bibinfo{year}{2012}).

\bibitem[{\citenamefont{e~Silva et~al.}(1997)\citenamefont{e~Silva, La~Rocca,
  and Bassani}}]{e1997spin}
\bibinfo{author}{\bibfnamefont{E.}~\bibnamefont{e~Silva}},
  \bibinfo{author}{\bibfnamefont{G.}~\bibnamefont{La~Rocca}}, \bibnamefont{and}
  \bibinfo{author}{\bibfnamefont{F.}~\bibnamefont{Bassani}},
  \bibinfo{journal}{Physical Review B} \textbf{\bibinfo{volume}{55}},
  \bibinfo{pages}{16293} (\bibinfo{year}{1997}).

\bibitem[{\citenamefont{Kanai and Shohno}(1963)}]{kanai1963dielectric}
\bibinfo{author}{\bibfnamefont{Y.}~\bibnamefont{Kanai}} \bibnamefont{and}
  \bibinfo{author}{\bibfnamefont{K.}~\bibnamefont{Shohno}},
  \bibinfo{journal}{Japanese Journal of Applied Physics}
  \textbf{\bibinfo{volume}{2}}, \bibinfo{pages}{6} (\bibinfo{year}{1963}).

\bibitem[{\citenamefont{Heremans et~al.}(2017)\citenamefont{Heremans, Cava, and
  Samarth}}]{heremans2017tetradymites}
\bibinfo{author}{\bibfnamefont{J.}~\bibnamefont{Heremans}},
  \bibinfo{author}{\bibfnamefont{R.}~\bibnamefont{Cava}}, \bibnamefont{and}
  \bibinfo{author}{\bibfnamefont{N.}~\bibnamefont{Samarth}},
  \bibinfo{journal}{Nature Reviews Materials} \textbf{\bibinfo{volume}{2}},
  \bibinfo{pages}{17049} (\bibinfo{year}{2017}).

\bibitem[{\citenamefont{Alves et~al.}(2013)\citenamefont{Alves, Neto, Scolfaro,
  Myers, and Borges}}]{alves2013lattice}
\bibinfo{author}{\bibfnamefont{H.}~\bibnamefont{Alves}},
  \bibinfo{author}{\bibfnamefont{A.}~\bibnamefont{Neto}},
  \bibinfo{author}{\bibfnamefont{L.}~\bibnamefont{Scolfaro}},
  \bibinfo{author}{\bibfnamefont{T.}~\bibnamefont{Myers}}, \bibnamefont{and}
  \bibinfo{author}{\bibfnamefont{P.}~\bibnamefont{Borges}},
  \bibinfo{journal}{Physical Review B} \textbf{\bibinfo{volume}{87}},
  \bibinfo{pages}{115204} (\bibinfo{year}{2013}).

\bibitem[{\citenamefont{Sengupta et~al.}(2017)\citenamefont{Sengupta, Tan,
  Klimeck, and Shi}}]{sengupta2017low}
\bibinfo{author}{\bibfnamefont{P.}~\bibnamefont{Sengupta}},
  \bibinfo{author}{\bibfnamefont{Y.}~\bibnamefont{Tan}},
  \bibinfo{author}{\bibfnamefont{G.}~\bibnamefont{Klimeck}}, \bibnamefont{and}
  \bibinfo{author}{\bibfnamefont{J.}~\bibnamefont{Shi}},
  \bibinfo{journal}{Journal of Physics: Condensed Matter}
  \textbf{\bibinfo{volume}{29}}, \bibinfo{pages}{405701}
  (\bibinfo{year}{2017}).

\bibitem[{\citenamefont{Bauer et~al.}(2012)\citenamefont{Bauer, Saitoh, and
  Van~Wees}}]{bauer2012spin}
\bibinfo{author}{\bibfnamefont{G.}~\bibnamefont{Bauer}},
  \bibinfo{author}{\bibfnamefont{E.}~\bibnamefont{Saitoh}}, \bibnamefont{and}
  \bibinfo{author}{\bibfnamefont{B.~J.} \bibnamefont{Van~Wees}},
  \bibinfo{journal}{Nature materials} \textbf{\bibinfo{volume}{11}},
  \bibinfo{pages}{391} (\bibinfo{year}{2012}).

\end{thebibliography}

\end{document}